\documentclass[prl,twocolumn]{revtex4}
\usepackage{graphicx}

\begin{document}

\title{Diffuse first-order phase transition in NaNbO$_3$:Gd}
\author{I. P. Raevski$^1$, S. I. Raevskaya$^1$, S. A. Prosandeev$^{1,2}$,
V. A. Shuvaeva$^{1,3}$, A. M. Glazer$^3$, M. S. Prosandeeva$^4$}
\affiliation{$^1$Rostov State University, Rostov on Don, Russia;
\\ $^2$National Institute of Standards and Technology,
Gaithersburg, MD; \\ $^3$Clarendon Laboratory, Oxford University,
Oxford, UK; \\ $^4$Institute of Theoretical and Experimental
Biophysics, Puschino, Russia} \maketitle

\begin{figure}
\resizebox{0.5\textwidth}{!} {\includegraphics{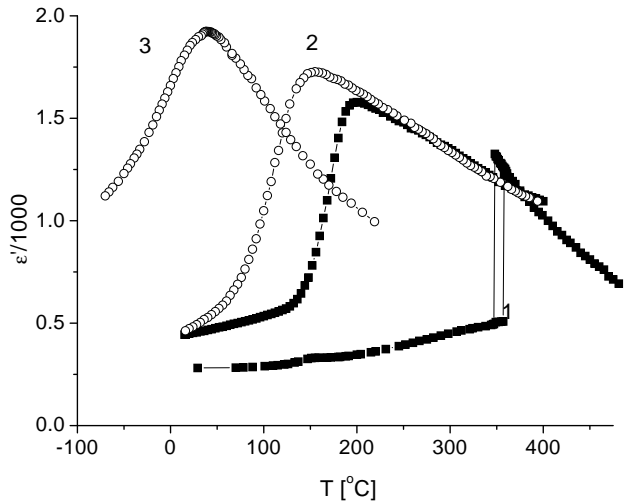}}
\caption{Dielectric permittivity in NNG0 (1), NNG9 (2) and NNG12
(3) measured at 100 KHz. The number after NNG shows the content of
Gd$_{1/3}$Nb$O_{3}$~ in mole percent.} \label{x}
\end{figure}

\begin{figure}
\resizebox{0.5\textwidth}{!} {\includegraphics{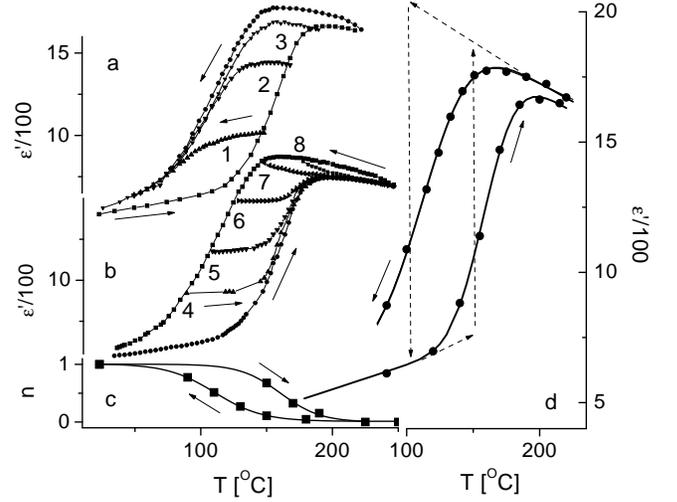}}
\caption{ Experimental data on dielectric permittivity of NNG9
obtained (a) for different annealing temperatures: $T_a\,[^\circ
C]=150$~ (1), 170 (2), 190 (3); (b) different cooling
temperatures: $T_{cool}\,[^\circ C]=90$~ (4), 110 (5), 130 (6),
150 (7), 180 (8); c) The LT-phase fraction, $n$, obtained from
experiment (squares); Solid lines are guides for the eye; (d)
Comparison of the fit of expression (\ref{eq1}) (solid line) to
experimental data (circles). Dotted lines show normal first-order
phase transition behaviour. The angles between the dashed and
solid lines characterize the degree of the diffuseness of the
phase transition.  }\label{hysteresis}
\end{figure}

Although diffuse phase transitions have been known for a long
time, in ferroelectric relaxors \cite{Smolenskii,Cross}, they have
not been studied as much in perovskite-based antiferroelectrics
\cite{Valant,Such,Raevski}. Despite this, the latter have some
obvious advantages in technical applications because of the
absence of toxic lead and strong frequency dispersion of
dielectric permittivity, $\varepsilon'$, inherent to relaxors.
NaNbO$_3$:Gd(NNG) crystals were grown by the flux method
\cite{Raevski}. At $x<0.1$~ $\varepsilon'$~ has hysteresis
(hereafter we mean thermal hysteresis of $\varepsilon'$), which
disappears at higher $x$~ (Fig. 1). Fig. 2 (a-b) shows
$\varepsilon'$~ obtained at heating of a NNG9 crystal up to
$T=T_a$~ with subsequent cooling (Fig. 2a), and cooling the
crystal down to $T_{cool}$~ with subsequent heating (Fig 2b). It
is seen that the hysteresis loop area depends on $T_a$~ and
$T_{cool}$~ strongly that implies that the diffuse first order
phase transition in NNG9 develops step by step in finite volumes.
As a result, the crystal's state becomes inhomogeneous. The
problem of describing dielectric permittivity in inhomogeneous
media is usually discussed in terms of the effective medium
approximation \cite{Bruggeman},  within which we obtained that the
fraction of the low-temperature (LT) phase is:

\begin{equation}
n(T)=\frac{[\varepsilon_2(T) - \varepsilon'(T)][\varepsilon_1(T) +
2\varepsilon'(T)]}{3\varepsilon'(T)[\varepsilon_2(T) -
\varepsilon_1(T)]}
\end{equation}
where $\varepsilon_1(T)$~ and $\varepsilon_2(T)$~ are the
dielectric permittivities in the  LT- and high-temperature (HT)
phases, respectively. All these quantities can be obtained from
the experiment performed. In agreement with our initial
assumption, the $n(T)$~dependence (Fig. 2c) is diffuse. One can
also use the hysteresis loop area as a (nonlinear) measure of
$n(T)$~ on cooling and $1-n(T)$~ on heating. Below, we will
discuss the shape of $n(T)$~ in more details.

Temperature-dependent optical studies of NNG have been carried out
by the rotating polarizer method, using the Metripol
(www.metripol.com) microscope system \cite{Glazer} and a precise
heating stage (Linkam HFS91). NNG crystals with $x=0.02-0.09$~
display distinctive changes of birefringence with temperature
similar to that in NaNbO$_3$.  At the phase transition point, a
spontaneous splitting into small regions (less than
$0.001\,\mathrm{mm}$~ in size) with diffuse boundaries occurs, and
the distribution of the birefringence image becomes very complex.
Afar of the phase transition temperature, the scale of this
non-uniformity increases up to $0.05\,\mathrm{mm}$.

The explanation of the experimental results obtained can be given
in the same way as for relaxors \cite{Smolenskii,Cross}. One can
introduce a distribution function for the Curie temperatures,
$f\left( {T_c } \right)$, which, in the simplest case, can be
described by a Gaussian function (this distribution is caused by
internal local fields and stresses introduced by impurities
\cite{Williamsb,Molak}):
\begin{eqnarray}
\label{eq1} \varepsilon (T) &=& \varepsilon_0 (T) + b(T)\int
{f(T_c) \theta (T - T_c) d T_c} = \nonumber \\ &=& \varepsilon_0 +
B \mathrm{erf}[(T - T_{c0})/\Gamma]
\end{eqnarray}
where $\varepsilon_0(T)$~ and $B(T)$~ are monotonic functions of
temperature, which, in the first approximation, can be given by
linear functions. Note that the Curie temperature of the
ferroelectric phase transition is sufficiently lower than the
temperature of the step. $\theta(x)$~ is a step function, which
equals 0 at $x<0$~ and 1 at $x>0$; erf(x) is the error function.
The fit (Fig. 2d) shows that the width of the distribution
function for NNG9 is about 27 K on heating and 35 K on cooling,
which is nearly comparable with the hysteresis width, 44 K. We
found that the distribution function width decreases with decrease
of the Gd content. The  results obtained allow one to suppose that
the disappearance of the hysteresis and the dramatic increase of
the $\varepsilon(T)$~ diffuseness observed experimentally
\cite{Raevski} in NNG12 (see Fig. 1), may be due to a crossover
between the widths of the $\varepsilon(T)$~ hysteresis and the
Curie temperature distribution function.

\begin{figure}
\resizebox{0.5\textwidth}{!} {\includegraphics{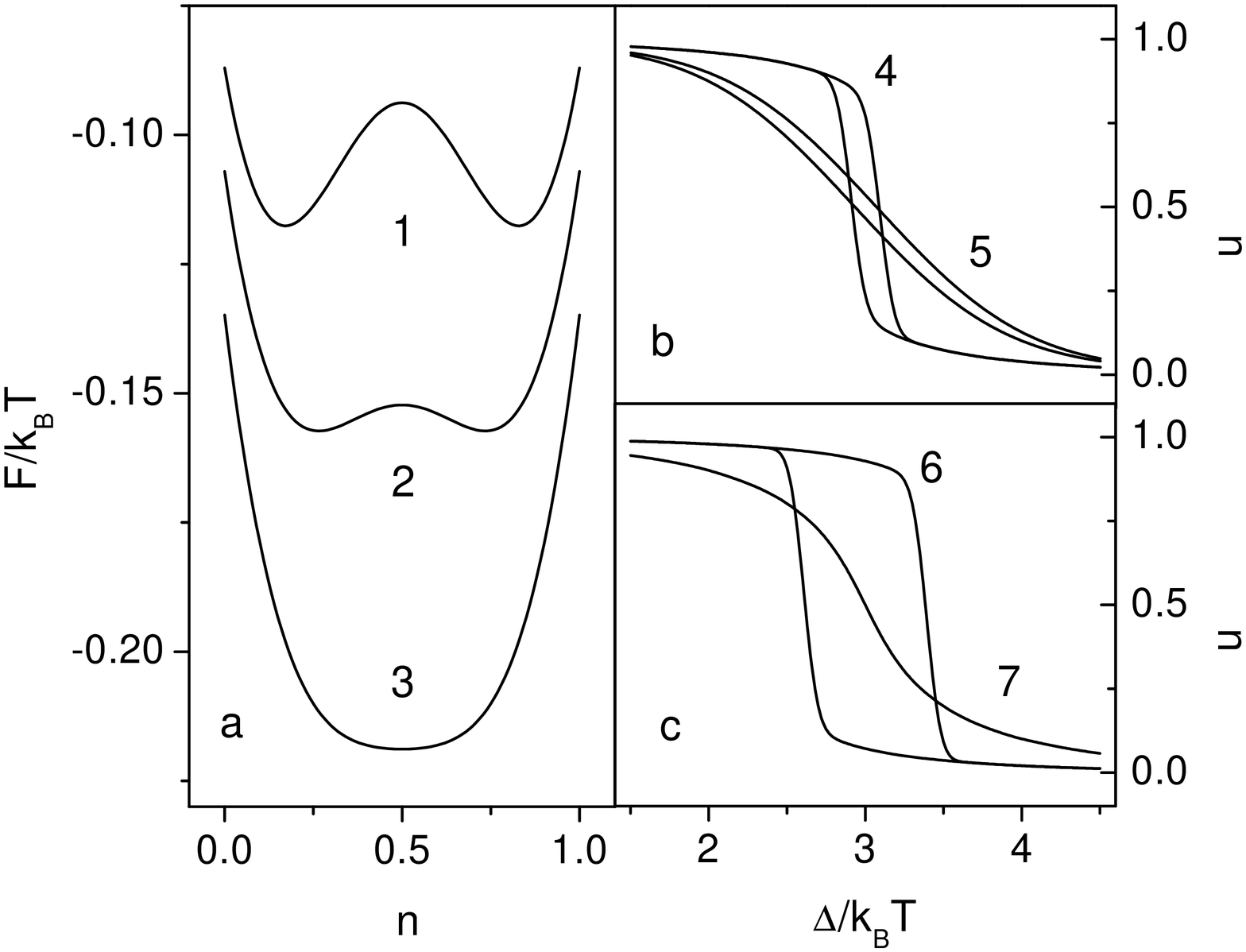}}
\caption{Theoretical curves for the model free energy (a): 1.
(2.4, 0.1); 2. (2.4, 1.0); 3. (2,4, 1.5); and order parameter $n$~
($b-c$): 4. (2.4, 0.1); 5. (2.4, 1.0); 6. (3.0, 0.1); 7. (1.5,
0.1) (The values in parentheses are $\gamma z/k_B T$~ and $a/k_B
T$). } \label{model}
\end{figure}

The diffuseness of a first order phase transition means coexisting
two phases. This must increase the surface tension energy. Below,
we discuss this surface tension contribution within a microscopic
approach, which is an extension of a lattice gas model \cite{H}:

\begin{equation}
\label{eq2} H = \Delta \sum {n_i + \gamma \sum {\left[ {n_i + n_j
- 2n_i n_j } \right]} }
\end{equation}
Here $\Delta$~ is a temperature dependent difference in chemical
potentials of the phases, $\Delta = k\left( {T - T_c } \right)$,
where we suggest $T_c$~ to be given randomly with a Gaussian
distribution function at $T = T_{c0}$; the eigen values of $n_{i}$
are 0 (HT phase) or 1 (LT phase); in the the second term we
introduce the interface energy due to surface tension (it is
positive when the nearest sites are different and zero when they
coincide). The summation runs over the sites each of which is
occupied by a solid either in the LT or HT phases. The summation
in the last term is over the nearest neighbors only. We have
derived the free energy (at a fixed value of $\Delta$), in the
mean field approximation: $F(\Delta) = - k_B T\ln \left( 1 + e^{h
/ k_B T} \right) + \gamma zn^2$~ where
$h=-\Delta-2z\gamma(0.5-n)$~ is a mean field and $n=<n_i>$~ is the
average occupation probability of the LT phase. $F$~ and $n$~ may
be averaged with a Gaussian function of $\Delta/k_B T$ having the
width $a$. We obtained that, the free energy has two minima at a
small $a$~ and at $\gamma > 2k_B T/z$~ where $z$~ is the number of
the nearest neighbors (Fig. 3a). Increasing the width of the
distribution function decreases the barrier height between the
states and decreases the inclination of the $n(\Delta)$~ curve
(the measure of the sharpness of the phase transition). The real
situation is, of course, more complex because of the random
distribution of barriers in space (the decrease of the barriers
may result in percolation of the new phase as it was found for
manganates \cite{Khomskii}), but the model does give a reasonable
explanation of observed behaviour: the phase transition is diffuse
in spite of the surface tension the role of which is suppressed
due to spreading the Curie temperatures.

\end{document}